\documentclass[10pt]{article}
\usepackage[
    paperwidth=210mm,
    paperheight=297mm,
    top=24mm,
    bottom=65mm,
    textwidth=135mm,
    headheight=5.5pt,
    headsep=5.6mm,
    marginparsep=5mm,
    marginparwidth=12mm,
    bindingoffset=0mm,
    footskip=10mm
]{geometry}
\usepackage{longtable}
\usepackage{ragged2e}
\usepackage{graphicx,psfrag,epsf}
\usepackage{enumerate}
\usepackage{url} 
\usepackage{float}
\usepackage{algorithm}
\usepackage{algorithmic}
\usepackage{amsmath}
\usepackage{amssymb}
\usepackage{subcaption}
\usepackage{natbib}
\usepackage{lineno}
\usepackage{dcolumn}
\usepackage{multirow}
\usepackage{comment}
\usepackage{booktabs}
\usepackage{colortbl}
\usepackage{rotating}
\usepackage{epstopdf}
\usepackage[table]{xcolor}  
\usepackage{siunitx}
\usepackage{setspace}

\usepackage[bottom]{footmisc}

\expandafter\def\expandafter\normalsize\expandafter{%
    \normalsize%
    \setlength\abovedisplayskip{5pt}%
    \setlength\belowdisplayskip{5pt}%
    \setlength\abovedisplayshortskip{5pt}%
    \setlength\belowdisplayshortskip{2pt}%
}

\newcolumntype{.}{D{.}{.}{-1}}
\newcolumntype{d}[1]{D{.}{.}{#1}}
\usepackage{hyperref}
\hypersetup{
	colorlinks=true,
	citecolor=blue,
}

\usepackage{rotating}
\usepackage{epsfig}%
\usepackage{epstopdf}

\def\btheta{\mbox{\boldmath $\theta$}}

\def\bf{\mbox{\boldmath $f$}}

\def\1{\mbox{1}}

\def\bveps{\mbox{\boldmath $\varepsilon$}}

\def\btheta{\mbox{\boldmath $\theta$}}

\def\bveps{\mbox{\boldmath $\varepsilon$}}

\def\btheta{\mbox{\boldmath $\theta$}}

\def\0{\mbox{\bf{0}}}

\def\0{\mbox{\bf{0}}}

\def\0{\mbox{\bf{0}}}

\newcommand{\nin}{\noindent}

\def\bkR{{\rm I\kern-.17em R}}

\def \1n{1\hskip -3pt \mbox{N}}

\newfont{\bbf}{cmbx12 scaled 1435}
\usepackage[normalem]{ulem}

\begin{document}
\thispagestyle{empty}

\renewcommand{\thefootnote}{\fnsymbol{footnote}}

\vspace*{0cm}

\begin{center}

{\fontsize{17}{17}\selectfont
\rmfamily
Nonfundamentalness or missing information ?\\
Evidence from causal-noncausal VARs in \\ macro-finance
}

\vspace{0.35in}

{\fontsize{12}{14}\selectfont
Lison Christiaens$^{1,2*}$,\quad
Julien Hambuckers$^{2,3\dagger}$\quad and\quad
Alain Hecq$^{1\dagger}$
}

\vspace{0.08in}

{\fontsize{12}{14}\selectfont
$^{1}$Maastricht University.\\
$^{2}$University of Liege - HEC Liege.\\
$^{3}$University of Göttingen.
}

\vspace{0.30in}

{\fontsize{12}{14}\selectfont
$^{*}$Corresponding author(s). E-mail(s):\\
\href{mailto:lison.christiaens@maastrichtuniversity.nl}
     {\textcolor{blue}{lison.christiaens@maastrichtuniversity.nl}};\\
Contributing authors:
\href{mailto:jhambuckers@uliege.be}
     {\textcolor{blue}{jhambuckers@uliege.be}};\\
\href{mailto:a.hecq@maastrichtuniversity.nl}
     {\textcolor{blue}{a.hecq@maastrichtuniversity.nl}};\\
$^{\dagger}$These authors contributed equally to this work.
}

\vspace{0.40in}

\begin{minipage}[t]{12cm}

\setlength{\baselineskip}{.115in}

\begin{center}
\textbf{Abstract}
\end{center}

In this paper, we study the presence of noncausal dynamics in standard macro-finance VAR models and ask whether they reflect omitted information available to economic agents but unobserved by the econometrician (nonfundamentalness), or pure forward looking behaviors. To that end, we introduce a factor-filtering mixed causal-noncausal VARX approach designed to account for common
macroeconomic information. We assess its performance in simulated settings, while showing also that the generalized covariance estimator correctly recovers causal and noncausal dynamics when using several lags. Empirically, we revisit the well-known Stock-Watson monetary policy (S)VAR and show that the noncausal components detected in the baseline specification largely disappear once common factors are filtered out. Finally, we compare impulse responses from the filtered and original data to assess the transmission of monetary policy shocks and show that filtering further removes the price puzzle.

\bigskip

\textbf{Keywords:} VAR models, nonfundamental shocks, GCov estimator, causal
and noncausal models, Non-Gaussianity

\bigskip

\textbf{JEL:} C32

\end{minipage}

\end{center}


\vfill

\noindent
\rule{\textwidth}{0.4pt}

\vspace{0.05in}

{\footnotesize
Earlier versions of this paper were presented at the 1st Workshop on
Noncausal Econometrics (2025), the CFE meeting (2025), the Time Series
Workshop (2026), the QFFE conference (2026), the IAAE conference (2026)
and at the ISF (2026). The authors would like to thank the participants
for their valuable comments and in particular C. Gourieroux, F. Giancaterini,
A. Neyazi and J. Jasiak. J. Hambuckers acknowledge the financial support
of the Fonds de la Recherche Scientifique - FNRS, grant number J009624,
under the project ``High-frequency tail risk dynamics in financial markets''.
}

\newpage

\renewcommand{\thefootnote}{\arabic{footnote}}

\section{Introduction}
Vector autoregressive (VAR) models have been widely used for decades in macro-finance to study dynamics in the economy, for forecasting and policy analysis (\citealp{sims1980macroeconomics}). However, structural shocks are typically assumed to be fundamental, meaning that they can be recovered from past values of forecast errors and therefore correspond to the shocks driving the data-generating process (DGP). This assumption is restrictive, as it constrains the model to a representation that may not be consistent with the true structure of the data. By contrast, in a nonfundamental representation, the structural shocks generating the economy do not coincide with - and cannot be recovered from - the prediction errors of a reduced-form VAR model \citep{KilianLutkepohl2017}. In the presence of nonfundamental components, standard VAR-based identification may therefore yield misspecified impulse response functions and misleading conclusions regarding the transmission of shocks. This possibility is typically not explicitly tested in the VAR literature. This is the objective of the present paper, which also seeks to evaluate whether nonfundamental components can influence empirical findings in macro-finance.

A common economic interpretation of nonfundamentalness is that economic agents observe more information than the econometrician. In this view, shocks may appear nonfundamental when the VAR model is too small to capture the relevant information driving the economy, or when economically relevant shocks are anticipated by agents. This insight has motivated a large literature on factor models (see \citealp{lippi1993dynamic,StockWatson2002PCA,Bernanke2005_FAVAR,FORNI2014fAVAR}), where large information sets are used to approximate agents’ information more closely.
Another route to address nonfundamentalness is more technical and consists in considering the presence of a non-invertible MA component in the VAR model. However, it is well-known in the time-series literature that Gaussian maximum-likelihood methods cannot discriminate between invertible and non-invertible models, which makes this approach more challenging to estimate \citep{lippi1994VAR}. Our goal in this paper is to reconcile these two views by combining (1) an approach that expands the information set available to the econometrician using common macroeconomic factors and (2) econometric tools that exploit non-Gaussianity to study nonfundamentalness. To this end, we introduce a mixed causal-noncausal VARX model to assess whether nonfundamentalness persists once the information set is expanded to include common macroeconomic factors. This paper is closely related to \cite{BEAUDRY2019221} who also tackle the issue of nonfundamentalness in economic SVAR model by developing a diagnostic of the nonfundamentalness severity.\footnote{Another related branch of the literature investigates nonfundamentalness in the scope of DSGE models, through models with anticipated shocks (see \citealp{forni2018_nonfund_DSGE,SOCCORSI201686}).} We also build on the sufficient-information test of \cite{FORNI2014fAVAR} to assess whether factor augmentation is sufficient to restore informational sufficiency and, when it is not, we introduce new tools to account for the remaining nonfundamentalness.

To do so, we rely on \textit{noncausal autoregressive models} rather than non-invertible MA representations. In these models, observed time series depend not only on their past, as in standard AR processes, but also on future values, which possibly captures nonfundamental dynamics arising from unobserved anticipations. Noncausal models can thus also be interpreted as capturing the mismatch of information between economic agents and the econometrician \citep{lanne2011noncausal}. This includes expectations of future endogenous variables, although their precise link to rational-expectations models has been less explored than in the case of non-invertible MA representations (see \citealp{BROCK20132710} for an example of forward-looking rational expectations systems and their nonfundamental representations).
An important feature of these noncausal models is that the linear projection on past values (e.g., VAR estimation by OLS) is not equal to the conditional expectation. More generally, causal and noncausal models, similarly to invertible and non-invertible MA processes, cannot be distinguished using only second-moment information.

To estimate the parameters of the VAR model, our main tool is the Portmanteau Generalized Covariance (GCov) estimator proposed by \cite{gourieroux2023generalized}. A specific challenge in the macro-financial context is the need to consider a multi-variable VAR model with more than one lag. These settings have received limited attention in the causal-noncausal literature, where most simulation evidence for the GCov estimator focuses on VAR(1) models. 
Because estimation becomes substantially more demanding as the parameter space expands, we first assess, in a Monte Carlo study, the finite-sample performance of the GCov estimator in a three-dimensional VAR(2) framework. In addition, we assess through simulations how the proposed factor-filtering VARX framework performs when noncausality arises from omitted information. This exercise serves as a proof-of-concept of our empirical study and provides guidance on the reliability of the estimator in realistic macroeconomic environments.

Empirically, we revisit the well-known study by \cite{stockwatson2001}, in which the authors develop a structural VAR model for U.S. inflation, unemployment, and the interest rate under alternative Taylor-rule specifications. Applying their approach to the original sample period (1960:I-2000:IV), we find the presence of several noncausal components that are not examined in the original study. To investigate whether these results reflect omitted macroeconomic information, we next filter the variables using common factors extracted from the FRED-QD database of \citet{McCrackenNg2021FREDQD},  and re-estimate the system. We find that once the first two macroeconomic factors are filtered out, the noncausal dimensions disappear in the Taylor-rule specifications. These results suggest that parts of the noncausal dynamics detected in the baseline VAR model reflect omitted aggregate information, and highlight the usefulness of noncausal models as diagnostic tools for uncovering hidden dynamics in small-scale systems. \cite{stockwatson2001} already discuss the potential role of omitted variables in canonical small VAR models and how this can bias the identification of monetary policy shocks. This concern motivated a large literature on factor models, pioneered by Stock and Watson themselves (see \citealp{StockWatson2002PCA} and subsequent work), which uses principal components to summarize information from large macroeconomic datasets. Nevertheless, this approach has not, to our knowledge, been explicitly applied to the three-variable VAR model in \cite{stockwatson2001}, which is what we tackle in this empirical application. Finally, we conclude by comparing impulse-response functions for monetary policy shocks across specifications and show that factor filtering further mitigates the price puzzle and soften the response of unemployment.

The rest of the paper is organized as follows. Section~\ref{sec:Noncausal} presents the representation and estimation of mixed causal-noncausal models and introduces a VARX framework to account for information captured by macroeconomic factors. Section~\ref{sec:montecarlo} provides Monte Carlo evidence on the finite-sample behavior of the GCov estimator in VAR(2) settings and evaluates the proposed VARX framework. Section~\ref{sec:empirical} revisits the Stock and Watson (2001) model under alternative Taylor-rule specifications, extends the analysis with factor filtering, and compares the resulting impulse responses. Section~\ref{sec:Conclusions} concludes.

\section{Causal-Noncausal Processes}
\label{sec:Noncausal}
This section first reviews the univariate mixed causal-noncausal framework before extending the discussion to the multivariate case, which forms the basis of our empirical analysis. 

\subsection{The Univariate Mixed Causal-Noncausal Model}
\label{sec:unicnc}

We begin by introducing the univariate causal-noncausal MAR($r,s$) specification
\begin{equation} \label{eq:MAR} 
    \Phi(L)\Psi(L^{-1})y_t = \epsilon_t,
\end{equation}
where the error term $\epsilon_t$ is non-Gaussian, independent, identically distributed (i.i.d.) and such that $E(|\epsilon_t|^{\delta}) < \infty$ for $\delta > 0$ (\cite{lanne2011noncausal}). The lag and lead polynomials $\Phi(L)$ and $\Psi(L^{-1})$ are of orders $r$ and $s$, respectively, and have all roots outside the unit circle. We do not include deterministic elements in \eqref{eq:MAR} in order to make notations easier but they can easily be added or alternatively one can partial them out from $y_t$. In applied research, causal-noncausal models were used to study various economic and financial variables (see inter alia \cite{cavaliere2020bootstrapping}, \cite{gourieroux2017local},
\cite{GourierouxHencicJAsiak2021_bubble},
\cite{jasiakhall}, 
\cite{hecq2021forecasting}, \cite{hecq2023predicting}, \cite{hencic2015noncausal}, \cite{lanne2013noncausal}, \cite{lof2017noncausality}).

We highlighted the existence of nonfundamental shocks in the introduction but most studies have considered the MAR($r,s$) model for its ability to capture complex nonlinear patterns, such as local trends, spikes and bubbles (see e.g. \cite{GHJNbubbles2025}). The MAR($r,s$) process \eqref{eq:MAR} admits a unique strictly stationary solution, which is a two-sided moving average of order infinity MA($\infty$):
\begin{equation*}
y_t = \sum_{h=-\infty}^{\infty} c_h \epsilon_{t-h},
\end{equation*}
in past, present and future shocks, with $c_0 = 1$  (\cite{breid1991maximum}). The coefficients $c_h$ are uniquely defined, provided that $(\epsilon_t)$ are non-Gaussian. When $y_t$ is purely noncausal (resp. causal), the coefficients $c_h$ are zero for all $h > 0$ (resp. $h < 0$). Thus, a purely causal process is determined only by the past and present shocks, while a purely noncausal process is influenced only by the present and future shocks. For $r=s=1$, we obtain the MAR($1,1$) process:
\begin{equation}
(1- \phi L)(1-\psi L^{-1}) y_t = \epsilon_t,
\end{equation}

\noindent with $|\psi|<1, |\phi|<1$, which is purely causal (resp. noncausal) if $\psi= 0$ (resp. $\phi= 0$). For each of these pure processes, the effects of a large $\epsilon_t$ are easily distinguished, as a large error leads to a (vertical) jump if $\psi = 0$ and $\phi>0$, and an explosive bubble with a (vertical) burst if $\psi > 0$  and $\phi=0$. 

An alternative univariate representation of the causal-noncausal process is
\begin{equation*}
    y_t = \varphi_1 y_{t-1} - \dots - \varphi_p y_{t-p} + \varepsilon_t,
    \label{eq:rootsIO}
\end{equation*}

\nin where $\varepsilon_t = -\frac{1}{\psi_s} \epsilon_t$ (\cite{brockwell1987stationary}). In this representation, the polynomial $\varPhi(L) = 1-\varphi_1 L - \cdots - \varphi_p L^p$, with $p=r+s$, has roots both inside and outside the unit circle. Specifically, the roots outside the unit circle correspond to the causal component of $y_t$, while the roots inside the unit circle indicate the noncausal component that captures bubbles and other nonlinear features. Note that the error $\varepsilon_t$ is not an innovation process because $\varepsilon_t$ is not independent of $y_{t-1}, y_{t-2},...$ and hence $E(y_t|y_{t-1}, y_{t-2}...)$ does not correspond to the linear projection  $LP(y_t|y_{t-1}, y_{t-2}...)$ because $E(\varepsilon_t|y_{t-1}, y_{t-2}...)$  $\neq 0$.\  

Consequently, the disturbance from the linear projection or an autoregressive model with a non correct allocation of roots is not an i.i.d. process, an observation that will lead to the development of Portmanteau type tests on residuals for identifying models. 

\subsection{The Multivariate Models}
\label{sec:GCov}

\indent \cite{lanne2013noncausal} propose the multivariate  VMAR($r$,$s$) model
\begin{equation}
\Phi (L) \Psi(L^{-1})Y_{t}=\epsilon_{t},
\end{equation}
where $\Phi (L)$ and $\Psi (L^{-1})$ are matrix 
polynomials of order $r$ and $s$ respectively and $Y_{t}$
is now a vector of $n$ time series. Similarly to the univariate case, both matrix polynomials are characterized by roots 
outside the unit circle
\begin{equation}
det (\Phi(z)) \neq 0\ \ \text{and}\ det (\Psi(z)) \neq 0,\ \ \text{for}\ \ |z| \leq 1,
\end{equation}
and $\epsilon_{t}$ is a sequence of $i.i.d.$ random 
non-Gaussian ($n \times 1$) vectors. However, unlike 
the univariate case, the model representation 
displayed in (3) presents two main differences. The first 
one is that the matrix multiplication being not 
commutative, $\Psi (L^{-1}) \Phi(L)Y_{t}=\epsilon_{t}$ 
provides different coefficient matrices than (3) for an observationally equivalent process. The second point about the multivariate noncausal representation (3) is that, unlike the univariate case, they cannot always be obtained from a VAR($n_1,n_2$) with roots inside and outside the unit circle (see \cite{lanne2013noncausal}, \cite{davis2020noncausal}, and \cite{gourieroux2017noncausal}). \\
\indent These two issues have motivated
\cite{davis2020noncausal}, 
\cite{gourieroux2017noncausal} or
\cite{gourieroux2023generalized} to directly start 
with a VAR($p$). 
Using this representation, the process is identified 
as purely causal if the $n \times p$ roots of the 
VAR($p$) lie outside the unit circle. 
In the opposite case, when the $n \times p$ roots 
are all inside the unit circle, the process is 
detected as purely noncausal. Finally, 
it is identified as mixed causal and noncausal when 
the roots lie both inside and outside the unit circle. We use the terms VMAR($r,s$) and VAR($n_1,n_2,p$) to distinguish between the model representations provided by Equation (3) and the VAR($p$) representation with $n_1$ roots outside and $n_2$ roots inside the unit circle in
\begin{equation}
Y_{t} = \Theta_1 Y_{t-1} + \Theta_2 Y_{t-2} + \dots + \Theta_p Y_{t-p} + u_t,
\label{eq:var(p)}
\end{equation}
where $det(\Theta(z)) = 0$ has $n_1$ roots outside and $n_2$ roots inside the unit circle, with $n_1 + n_2 = n \times p$.  Although the VAR($p$) process is a linear model, it displays nonlinear features when one or more of its roots are situated inside the unit circle. As a result, in such instances, the right-hand side first part of the equation ceases to represent the conditional expectation of a linear model. The error term $u_t$ must still meet the requirements of being $i.i.d.$ as well as non-Gaussian.  It will not be $i.i.d.$ using a linear projection in the presence of a noncausal component, namely when  $n_2 > 0$.

To prove the existence of a stationary solution of (5) and derive its two-sided moving average representation, it is necessary to introduce the \textit{Representation Theorem} proposed by \cite{gourieroux2017noncausal}.

\textbf{Representation theorem} (\textit{from 
\cite{gourieroux2017noncausal}}):

Let us consider a mixed causal-noncausal VAR($p$) model as in equation~(5), that can be written as mixed causal-noncausal VAR(1) model by using the companion form as follows (\cite{gourieroux2017noncausal}):
\begin{equation*}
    X_t= \Psi X_{t-1} + \xi_t,
\end{equation*}
where $X_t=[Y_t, Y_{t-1}, \dots , Y_{t-p+1} ]^{\prime}$, $\xi_t=[u_t, 0, 0, \dots, 0]$. 

There exists an invertible $(np \times np)$ real matrix $B$, and two square real matrices $J_{1}$ and $J_{2}$ of dimensions $(n_{1}\times n_{1})$ and $(n_{2}\times n_{2})$, with $n_{1}+n_{2}=np$, such that all eigenvalues of $J_{1}$ (resp. $J_{2}$) correspond to those of $\Psi$ with modulus strictly less (resp. greater) than one:
\begin{equation*}
    \Psi = B
    \begin{bmatrix}
        J_1 & 0\\
        0 & J_2
    \end{bmatrix}
    B^{-1}.
\end{equation*}
As a consequence, we have:
\begin{gather}
X_t = B_{1}X_{1,t}^{*} + B_{2} X_{2,t}^{*} \\
X_{1,t}^{*} = J_{1}X_{1,t-1}^{*} + \xi_{1,t}^{*}, 
\qquad
X_{2,t}^{*} = J_{2}X_{2,t+1}^{*} + \xi_{2,t}^{*} \\
X_{1,t}^{*} = B^{1}X_{t}, \qquad X_{2,t}^{*} = B^{2}X_{t} \\
\xi_{1,t}^{*} = B^{1}\xi_{t}, \qquad \xi_{2,t}^{*} = B^{2}\xi_{t}
\end{gather}
where $[B_{1},B_{2}]=B$ and $[B^{1\prime},B^{2\prime}]^{\prime}=B^{-1}$. 
The matrix $J_2$ must be invertible, consequently any zero eigenvalues will be accommodated within matrix $J_1$. This implies that while $n_1$ can include roots equal to 0, $n_2$ cannot. 

Note that in a VAR(1) model, using equation (8), we can observe that $X_{1,t}^*$ and $X_{2,t}^*$ represent purely causal and noncausal processes, respectively. Therefore, we can interpret these two components as the causal and noncausal components of the process $X_t$. Additionally, when $p \geq 2$, the causal and noncausal components are functions of the current and lagged values of $Y_t$, since $X^*_{1,t}=B^1X_t=\sum_{h=0}^{p-1}B^1_h Y_{t-h}$ and $X^*_{2,t}=B^2X_t=\sum_{h=0}^{p-1}B^2_h Y_{t-h}$.

The VAR($n_1, n_2, p$) specification represents a more general framework than the VMAR($r,s$) since it does not rely on a multiplicative structure and hence the number of roots related to the causal and noncausal components in (5) does not necessarily have to be a multiple of $n$. In the context of univariate time series, it is always possible to convert a process from representation (5) to (3) and vice versa. However, in the multivariate setting, this is not straightforward. Giancaterini (2023) proved the conditions between VAR($n_1, n_2, p$) and VMAR($r,s$) models. The results we are interested in are summarized in the following theorem.

\textbf{Theorem 1:} (Giancaterini, 2023) Let $Y_t$ be an $n$-dimensional time series generated by the stationary process expressed as (5), i.e., a VAR($n_1, n_2, p$) process. Then, $Y_t$ has a VMAR($r,s$) specification if and only if:
\begin{equation}
    \frac{n_1}{n} \ \in \ \mathbb{N} \ \ and \ \ \frac{n_2}{n} \ \in \ \mathbb{N}.
\end{equation}\label{theorem_n1/n}
This is the generalization of the example in \cite{lanne2013noncausal} for the VAR(1) model.

\subsection{A VARX model for factor filtering}
\label{sec:VARX}

Building on the mixed causal-noncausal framework developed above, we now extend the analysis by allowing for exogenous variables within a VARX specification. \cite{HecqISST} introduce a MARX model that incorporates strictly exogenous regressors into a mixed causal-noncausal setting. The idea is to add variables such as the output gap to a mixed causal-noncausal model for inflation. The motivation in this paper is more ``fundamental". Indeed, we have argued that the presence of a noncausal component, and hence the existence of nonfundamental shocks in a VAR model, may be attributed to the mismatch between the information of econometricians and economic agents. Our application, revisiting the three-dimensional study of \cite{stockwatson2001}, is a good example of a small-dimensional system.
We propose a new framework in which we expand the VAR($n_1, n_2, p$) model with additional exogenous variables such as:
\begin{equation}
Y_{t} = \Theta_1 Y_{t-1}  + \dots + \Theta_p Y_{t-p} + \Gamma F_t + u_t.
\end{equation}
Several key economic variables can be considered for $F_t$, such as oil prices. Our objective, however, is to incorporate proxies for the broader information set available to economic agents. In the empirical application, we therefore use common factors extracted from a large panel of stationary macroeconomic time series drawn from the FRED-QD database.

Our approach differs from the univariate MARX model (\cite{HecqISST}) in two aspects. First, we are not interested in the coefficients $\Gamma$ per se. We include the factors to partial out the small VAR model by a large amount of information to detect whether the noncausality that could be detected in a small setting is still present. Second, the MARX uses the multiplicative form, a strategy that could lead to the multivariate VARX model such that 
\begin{equation}
\Phi (L) \Psi(L^{-1})Y_{t}= \Gamma F_t+ \epsilon_{t}.
\end{equation}

Let $F_{j,t}$ denote the $j$th factor extracted from the FRED-QD dataset.
For each observable of $Y_t$, we implement the VARX model introduced by allowing for contemporaneous and strictly exogenous factors $F_{j,t}$. Rather than estimating the full VARX system jointly, we rely on a partialling-out approach that is equivalent for retrieving the autoregressive parameters of interest. Specifically, we estimate the projection
\begin{equation}
Y_{t-p} = \alpha_p + \sum_{j=1}^{J} \beta_{p,j} F_{j,t} + \tilde{Y}_{t-p}
\label{eq:varxp}
\end{equation}
where $J$ denotes the number of factors included and $p$ the lag number. By the Frisch-Waugh-Lovell theorem (see \citealp{greene2018econometric}), estimating a VAR model on the residual component $\tilde{Y}_t$ is equivalent to estimating a VARX model with contemporaneous exogenous regressors $F_{j,t}$ and focusing on the dynamics of $Y_t$ conditional on these factors. The residual $\tilde{Y}_t$ therefore represents the component of $Y_t$ orthogonal to the selected information set $F_t$. The same projection is applied to the lagged variables entering the VAR model, ensuring that both contemporaneous observations and lagged regressors are consistently filtered out from common macroeconomic components. Further details are provided in the empirical application.

\subsection{The GCov Estimator}
\label{sec:GCov}

\indent To estimate the proposed causal-noncausal model, we rely on the semi-parametric  Generalized Covariance (GCov) estimator (\cite{gourieroux2023generalized}). This estimator does not require any distributional assumptions on the errors, other than non-Gaussianity for the identification of parameters. It is a one-step estimator that is consistent, asymptotically normally distributed, and semi-parametrically efficient. The GCov minimizes a portmanteau-type objective function involving nonlinear autocovariances, that is, the autocovariances of nonlinear transformations of model errors $a\left(\bveps_t\right)= \left[ a_1\left(\bveps_{t}\right)^{\prime}, \dots , a_K\left(\bveps_t\right)^{\prime}\right]$ that satisfy the regularity conditions given in \cite{gourieroux2023generalized}. This increases the dimension of the process from 1 to $K$. Let us denote by $\hat{\Gamma}^a\left(h; \btheta\right), h=1,...,H$ the autocovariance matrices at lags $h=0,...,H$, with $\hat{\Gamma}^a\left(0; \btheta\right)$ representing the variance, and $\btheta$ the vector of autoregressive parameters. Then,
the GCov estimator of the parameter $\hat{\btheta}_{T}$ minimizes the following objective function:
\begin{equation}
    \hat{\btheta}_{T}= \underset{\btheta}{\mathrm{argmin}} \sum_{h=1}^{H} Tr \left[ \hat{\Gamma}^a\left(h; \btheta\right) \hat{\Gamma}^a\left(0; \btheta\right)^{-1}\hat{\Gamma}^a\left(h; \btheta\right)' \hat{\Gamma}^a\left(0; \btheta\right)^{-1} \right],
    \label{eq:GCov22}
\end{equation}
\nin where $Tr$ denotes the trace of a matrix and we denote $L_T( \hat{\btheta}_T,H)$ the value of the function at its minimum.\footnote{Alternatively, when the number of transformations $K$ is large, we can replace in the above formula $\hat{\Gamma}^a\left(0; \btheta\right)$ with  $diag\left(\hat{\Gamma}_d(0; \btheta)\right)$, which is the variance matrix containing only the diagonal elements of $\hat{\Gamma}_a(0; \theta)$. This latter version of the GCov estimator is no longer semi parametrically efficient, as it is not optimally weighted (see \cite{gourieroux2023generalized}, \cite{cubadda2011testing}). An alternative to the diagonal GCov estimator is the regularized GCov method proposed by \cite{giancaterini2025regularized}.}

The choice of an informative set of transformations ($a_{k}, k=1,..., K$) depends on the specific series under investigation. For example, in financial applications, linear and quadratic functions can be selected, such as $a_1(\varepsilon_t) =\varepsilon_{t}$, $a_2(\varepsilon_t) =  \varepsilon_{t}^2$. 

In the VARX model, we partial out $Y_t$, $Y_{t-1},...Y_{t-p}$ by $F_t$ and we implement the GCov estimator on the residuals. The factors being stationary, that does not affect the consistency of the GCov estimator.

\section{Monte Carlo Results}
\label{sec:montecarlo}
Only a few studies have investigated the finite sample behavior of the GCov estimator. In general, they focus on the robustness of the estimator for different choices of transformations and initial conditions for the VAR parameters (see \cite{cubadda2023optimization} and \cite{giancaterini2025regularized}). Moreover, the VAR(1) model is usually considered. Therefore, in this section, we provide guidelines in a VAR(2) context, since more than one lag is usually required in macro-finance applications. In addition, we test how the VARX setting can reconsider the mismatch of information in a novel nonfundamentalness specification.

\subsection{Performance of the GCov Estimator}

There are two main reasons for looking at the GCov estimator in a VAR(2) model. First, the empirical application of the next section necessitates a lag length higher than a VAR(1) model. Second, the reliability of the GCov estimator may depend on the choice of starting values used in the optimization procedure. \cite{cubadda2023optimization} show that the GCov estimator can lack robustness with respect to initialization and propose a Simulated Annealing (SA) algorithm to improve convergence toward the global optimum. However, implementing such a strategy becomes considerably more computationally demanding in a VAR(2) setting. For instance, in a three-variable system, the autoregressive block alone involves 18 parameters, significantly increasing the dimension of the optimization problem. Given this limitation, we assess the finite-sample performance of the GCov estimator in a VAR(2) model using standard starting values in the literature.

We compare the results for the VAR(1) and VAR(2) models in Tables~\ref{tab:causal_dim_results} and \ref{tab:causal_var2}, respectively. The VAR(1) model is also estimated to provide a benchmark under identical DGPs and simulation settings. All Monte Carlo experiments consider $T = \{250, 500, 1000\}$ observations and $N = 1000$ replications. The error term is assumed to be distributed according to a multivariate Student's $t$-distribution with $\nu = 4$ degrees of freedom, and a diagonal variance-covariance matrix, i.e., $\Sigma = \nu /(\nu - 2)I_{3}$. Although the fourth moments do not exist when $\nu = 4$, the identification of the models is preserved as it only requires consistency. We have considered many choices of the transformations and the autocovariance lags but we only report the outcomes obtained from the best simulations. In particular, the nonlinear transformations $\{e_t, e_t^2, e_t^3, e_{ti}e_{tj}\}$ and autocovariances up to $H=6$.\footnote{Other results are available upon request.} These choices are maintained as the baseline specification throughout the remainder of the paper, including in the empirical application.

The main results are the following. For the VAR(1) specification, we observe a high frequency with which the correct specification VAR(2,1,1) is obtained. When the true parameters are taken to initialize the process, the frequencies range from 82.3\% to 97.6\%; they go from 72.3\% to 93.1\% if OLS starting values are used. This decrease has also been observed in \cite{cubadda2023optimization}. The averages of the eigenvalues (inverse of the roots) are very close to those in the DGP. We also provide, in the online appendix, Figure~E.1 that displays the distribution, over 1,000 replications, of the nine estimated parameters for the VAR(1) model. Although some replications yield parameter estimates far from the true values, the distortions are less pronounced than those reported in \cite{cubadda2023optimization}. This result also highlights the potential influence of the chosen DGP parameters on the finite-sample results.
  
We then extend the simulations to the VAR(2) case and consider several DGPs for the VAR($n_1, n_2, 2$) specification. These are a VAR($6, 0, 2$), a VAR($4, 2, 2$), a VAR($3, 3, 2$) and a VAR($2, 4, 2$). The VAR($6, 0, 2$) is equivalent to a purely VAR(2) causal model. Note that only the mixed VAR($3, 3, 2$) (and of course the purely VAR($6, 0, 2$)) satisfies the condition of Theorem\hyperref[theorem_n1/n]{~1}. Important conclusions emerge from Table~\ref{tab:causal_var2}. We first observe that the frequency with which the correct model is selected increases with the sample size, illustrating the consistency of the GCov estimator. This suggests that the method may be particularly well-suited to financial time series, which are typically available at higher frequencies and also tend to exhibit stronger non-Gaussian features than standard macroeconomic data. The consistency is also emphasized when looking at the decrease of RMSE and MAE with the sample size.  Second, there is sometimes a large difference in the frequencies obtained between simulations starting from the true initial values and those obtained from OLS in a VAR(2) model. Obviously, the two types of initial conditions give close results when $n_1$ is large. The extreme case is DGP A, that is actually a purely causal VAR(2) model. Finally, misclassification amounts at most to an over- or underestimation of one noncausal root (typically due to the presence of complex conjugate roots that come in pairs). In particular, for DGPs B, C and D, we tend to underestimate the number of noncausal dimensions\footnote{In the online Appendix, Section B, we also study the frequencies with which we obtain a VAR(3,3,2) specification when one of the roots gets closer to the unit root. One observes that the frequencies of choosing the correct model are mainly affected when we have a root very close to 1, e.g. 0.99. Still, we obtain the VAR(3,3,2) specification in 60\% of the cases.}.  

In summary, the Monte Carlo results indicate that the GCov estimator can be reliably applied to our three-dimensional VAR(2) framework. While small-sample distortions are more pronounced for $T=250$, the correct causal dimension is recovered with high frequency, and performance improves quickly with the sample size. 

\begin{table}
\centering
\footnotesize
\caption{Classification frequencies and median eigenvalues by causal dimension}
\label{tab:causal_dim_results}

\begin{tabular*}{\textwidth}{@{\extracolsep{\fill}} l c c c c c c c c c}
\toprule
Causal dim.
& 0 & 1 & 2 & 3
& $\lambda_1$ & $\lambda_2$ & $\lambda_3$
& MAE & RMSE \\
\midrule
\multicolumn{10}{l}{\textbf{DGP:} $n_1 = 2$, $\lambda = (1.3,\;0.7,\;0.3)$} \\[2pt]

\% True, $T=250$
& 0.2 & 11.1 & \textbf{82.3} & 6.4
& 1.3498 & 0.7165 & 0.5032
& 0.1544 & 0.1247 \\

\% True, $T=500$
& 0 & 4.3 & \textbf{92.0} & 3.7
& 1.3331 & 0.6993 & 0.5063
& 0.0984 & 0.1218 \\

\% True, $T=1000$
& 0 & 0.9 & \textbf{97.6} & 1.5
& 1.3148 & 0.6988 & 0.5026
& 0.0632 & 0.0780 \\

\addlinespace[4pt]
\% OLS, $T=250$
& 0 & 11.6 & \textbf{72.3} & 16.1
& 1.2900 & 0.7113 & 0.4965
& 0.1445 & 0.1795 \\

\% OLS, $T=500$
& 0 & 4.9 & \textbf{85.0} & 10.1
& 1.3156 & 0.7003 & 0.5068
& 0.0991 & 0.1236 \\

\% OLS, $T=1000$
& 0 & 1.2 & \textbf{93.1} & 5.7
& 1.2941 & 0.6932 & 0.4952
& 0.0648 & 0.0805 \\
\bottomrule
\end{tabular*}

\vspace{3pt}
\begin{minipage}{\textwidth}
\footnotesize
\justifying
\emph{Notes:}
This table reports classification frequencies for the number of causal dimensions identified by the GCov estimator in a VAR(1) model. The true number of causal roots is $n_1=2$ (bold).
GCov is estimated using the nonlinear transformations $\{e_t, e_t^2, e_t^3, e_{ti}e_{tj}\}$ and autocovariances up to $H=6$, based on 1,000 Monte Carlo replications.
Median eigenvalues are reported as $\lambda_1$, $\lambda_2$, and $\lambda_3$.
MAE denotes the Frobenius average deviation,
$\text{MAE}=\frac{1}{q}\sum_{i=1}^p|\Theta_i^{\text{true}}-\Theta_i^{\text{hat}}|$, with $q=9$.
\end{minipage}

\end{table}

\begin{table}[h]
\centering
\footnotesize
\caption{Classification frequencies by number of causal dimensions estimated with GCov}
\label{tab:causal_var2}

\begin{tabular*}{\textwidth}{@{\extracolsep{\fill}} lcccccccSS}
\toprule
Causal dim.
& 0 & 1 & 2 & 3 & 4 & 5 & 6
& \multicolumn{1}{c}{MAE}
& \multicolumn{1}{c}{RMSE} \\
\midrule

\multicolumn{10}{l}{\textbf{DGP A (purely causal):} $n_1=6$, $\lambda=(0.85,\;0.75,\;0.6,\;0.5,\;0.4,\;0.3)$} \\[2pt]
\% True, $T=250$   & 0 & 0 & 0.4 & 4 & 20 & 40.7 & \textbf{34.9}   & 0.1739 & 0.2192 \\
\% True, $T=500$   & 0 & 0 & 0   & 0.7 & 8.9    & 34.9 & \textbf{55.5} & 0.0964 &  0.1213 \\
\% True, $T=1000$  & 0 & 0 & 0   & 0.2   & 2.3  & 21   & \textbf{76.5} & 0.0455 & 0.0577 \\
\addlinespace[4pt]
\% OLS, $T=250$    & 0 & 0 & 0.4   & 3.8 & 20.2 & 40 & \textbf{35.6} & 0.1727 & 0.2163 \\
\% OLS, $T=500$    & 0 & 0 & 0   & 0.6 & 9  & 35.2 & \textbf{55.2} & 0.0988 & 0.1240 \\
\% OLS, $T=1000$   & 0 & 0 & 0   & 0.2   & 2.3  & 21 & \textbf{76.5}   & 0.0452 &  0.0573 \\

\midrule
\multicolumn{10}{l}{\textbf{DGP B (mixed):} $n_1=4$, $\lambda=(1.3,\;1.1,\;0.85,\;0.7,\;0.5,\;0.3)$} \\[2pt]
\% True, $T=250$   & 0 & 0.1 & 1   & 17.5 & \textbf{63.7} & 17.2 & 0.5 & 0.1783 & 0.2268 \\
\% True, $T=500$   & 0 & 0   & 0.1 & 7.8  & \textbf{78.1} & 13.6 & 0.4 & 0.0966 & 0.1194 \\
\% True, $T=1000$  & 0 & 0   & 0   & 3.1  & \textbf{90.3} & 6.6  & 0   & 0.0633 & 0.0776 \\
\addlinespace[4pt]
\% OLS, $T=250$    & 0 & 0.1 & 2.3   & 19.8 & \textbf{50} & 25.2 & 2.6 & 0.2140 &  0.2749 \\
\% OLS, $T=500$    & 0 & 0   & 0.7 & 13.6  & \textbf{60.6} & 23.8 & 1.3 & 0.1323 & 0.1659 \\
\% OLS, $T=1000$   & 0 & 0   & 0.1   & 5.9  & \textbf{78.1}   & 15.7  & 0.2   & 0.0837 &  0.1031 \\

\midrule
\multicolumn{10}{l}{\textbf{DGP C (mixed):} $n_1=3$, $\lambda=(1.5,\;1.4,\;1.3,\;0.7,\;0.5,\;0.3)$} \\[2pt]
\% True, $T=250$   & 0 & 0.2 & 7.8 & \textbf{70.8} & 20.2 & 1   & 0   & 0.4441 & 0.5824 \\
\% True, $T=500$   & 0 & 0   & 3.1 & \textbf{83.5}  & 13   & 0.4 & 0   & 0.2708 & 0.3469 \\
\% True, $T=1000$  & 0 & 0   & 0.2 & \textbf{95.6}  & 4.2  & 0   & 0   & 0.1371 & 0.1718 \\
\addlinespace[4pt]
\% OLS, $T=250$    & 0 & 0.1 & 6.4 & \textbf{36.8}  & 43.1 & 12.4& 1.2 & 0.3457 & 0.4486 \\
\% OLS, $T=500$    & 0 & 0   & 1.3 & \textbf{53.3}  & 38.8 & 5.9 & 0.7 & 0.1760 & 0.3080 \\
\% OLS, $T=1000$   & 0 & 0   & 0.3 & \textbf{74.9}  & 23.1 & 1.7 & 0   & 0.1399 & 0.1764 \\

\midrule
\multicolumn{10}{l}{\textbf{DGP D (mixed):} $n_1=2$, $\lambda=(1.4,\;1.3,\;1.2,\;1.1,\;0.5,\;0.3)$} \\[2pt]
\% True, $T=250$   & 0 & 0.5 & \textbf{39.3} & 43.8 & 15.1 & 1.3 & 0   & 0.2577 & 0.3236 \\
\% True, $T=500$   & 0 & 0.2 & \textbf{55.4} & 35.5 & 8.8  & 0.1 & 0   & 0.1677 & 0.2107 \\
\% True, $T=1000$  & 0 & 0   & \textbf{73.9} & 21.9 & 4    & 0.2 & 0   & 0.1053 & 0.1322 \\
\addlinespace[4pt]
\% OLS, $T=250$    & 0 & 0.7 & \textbf{27.9} & 37.5 & 24.9 & 8.3 & 0.7 & 0.2728 & 0.3438 \\
\% OLS, $T=500$    & 0 & 0.2 & \textbf{47.5} & 35.3 & 14.2 & 2.8 & 0   & 0.1750 & 0.2216 \\
\% OLS, $T=1000$   & 0 & 0   & \textbf{70.9} & 21.7 & 6.9  & 0.5 & 0   & 0.1010 & 0.1388 \\
\bottomrule
\end{tabular*}

\vspace{3pt}
\begin{minipage}{\textwidth}
\footnotesize
\justifying
\emph{Notes:} GCov is estimated using the nonlinear transformations $\{e_t, e_t^2, e_t^3, e_{ti}e_{tj}\}$ and autocovariances up to $H=6$, based on 1{,}000 Monte Carlo replications.
\end{minipage}
\end{table}

\subsection{Assessing nonfundamentalness with VARX}

We now assess whether the noncausal dimensions detected by GCov may arise from an informational mismatch rather than from a truly noncausal data-generating process. We propose the following benchmark, a purely causal factor-augmented VAR(6,0,2) specification,
\begin{equation}
    Y_t = \Theta_1 Y_{t-1} + \Theta_2 Y_{t-2} + \Gamma F_t + u_t,
\label{eq:varx2}
\end{equation}
where $Y_t$ is the $3 \times 1$ vector of macroeconomic variables, $F_t$ denotes a set of latent information factors, and $u_t$ is an i.i.d. non-Gaussian innovation. The coefficient matrices $(\Theta_1,\Theta_2)$ are chosen so that the companion roots imply a purely causal system. The factor process is taken directly from the empirical factors used in the application. We consider three factors in the DGP as a trade-off between capturing sufficient information and limiting model complexity. Empirical evidence (e.g., \cite{McCrackenNg2021FREDQD}) shows that the marginal explanatory power declines quickly after the first few factors, a pattern also observed in our application.

The central hypothesis is that noncausality should not be detected when the econometrician conditions on the same information set as the one of the true data-generating process. By contrast, when relevant factors are omitted from estimation, GCov may yield a nonzero noncausal dimension, reflecting informational mismatch.
More formally, the Monte Carlo design is built around the following predictions. First, when GCov is applied to the correctly specified factor-filtered system, the estimated noncausal dimension should satisfy $ n_2 = 0$ with high probability. Second, when the same simulated data are estimated with a small-information VAR model that omits the relevant factors, the probability of finding $ n_2 > 0$ should increase. Third, when only a subset of the relevant factors is included, the frequency with which GCov detects noncausality should decline as the information set becomes richer.

The simulation proceeds in three steps. First, a causal VAR(6,0,2) model with factors is calibrated using parameter values consistent with the empirical application, and samples are simulated from this benchmark model. Second, for each simulated sample, GCov is estimated under alternative information sets: (i) without factors, (ii) with a restricted subset of factors, and (iii) with the full factor set. Third, for each specification, we record the estimated coefficient matrices, companion roots, and the implied causal and noncausal dimensions.

\paragraph{Factor filtering:}
Each regressor entering the VAR(6,0,2) model is individually residualized with respect to the contemporaneous factor vector. For each horizon $h = 0,1,2$, we run the projection
\begin{equation}
    Y_{t-h} = a_h + B_h F_t + \tilde{Y}_{t-h},
\end{equation}
where $a_j$ is a vector of constants and $B_j$ is the matrix of factor loadings estimated from the projection of $Y_{t-h}$ on $F_t$. The residual $\tilde{Y}_{t-h}$ represents the component of $Y_{t-h}$ that is orthogonal to the contemporaneous factor space. The VAR(2) model is then estimated using the filtered regressors,
\begin{equation}
    \tilde{Y}_t = \Theta_1 \tilde{Y}_{t-1} + \Theta_2 \tilde{Y}_{t-2} + \varepsilon_t.
\end{equation}
This approach ensures that both the dependent variable and its lagged regressors are orthogonal to the contemporaneous factor space.

We additionally test our filtering procedure by taking the true values of $\Gamma, \Theta_1, \Theta_2$ to compute the residuals $\hat\varepsilon_t$ of the corresponding VAR(2) model, following $\tilde{Y}_t - \Gamma^{\text{DGP}} F_t
= \Theta_1^{\text{DGP}} \tilde{Y}_{t-1}
+ \Theta_2^{\text{DGP}} \tilde{Y}_{t-2}
+ \varepsilon_t$. This allows us to see the number of noncausal dimensions found without the bias from estimating the matrix of loadings and coefficients in Equation~\ref{eq:varx2}.

\begin{table}
\centering
\footnotesize
\caption{Classification frequencies for the estimated number of causal dimensions (T=1000)}
\label{tab:causal_varX2}

\begin{tabular*}{\textwidth}{@{\extracolsep{\fill}} lcccccccSS}
\toprule
 & \multicolumn{7}{c}{Estimated causal dimension ($n_1$)} & \multicolumn{1}{c}{MAE} & \multicolumn{1}{c}{RMSE} \\
  & 0 & 1 & 2 & 3 & 4 & 5 & 6 &  &  \\
\midrule

\multicolumn{10}{l}{\textbf{Purely causal DGP simulated with 3 factors}, estimated using:} \\
GCov (3 factors)  & 0 & 0 & 0 & 0.3 & 5.1 & 24.5 & \textbf{70.1} &  0.0589 & 0.0760 \\
GCov (2 factors) & 0 & 0 & 0 & 0.7 & 4.5 & 26.9 & \textbf{67.9} & 0.0587 & 0.0758 \\
GCov (1 factor)  & 0 & 0 & 0 & 0.2 & 6.1 & 25.1 & \textbf{68.6} & 0.0785 & 0.0982 \\
GCov (0 factor)  & 0 & 0 & 0 & 0.8 & 13 & 34 & \textbf{52.2} & 0.1261 & 0.1598 \\
\midrule

\multicolumn{10}{l}{\textbf{Purely causal DGP simulated with 3 factors}, estimated using true $\Gamma, \Theta_1, \Theta_2$ and:}\\
GCov (3 factors)  & 0 & 0 & 0 & 0.7 & 5.4 & 17.9 & \textbf{76} & 0.0620 & 0.0799 \\
GCov (2 factors) & 0 & 0 & 0.1 & 0.6 & 5.1 & 23.6 & \textbf{70.6} & 0.0594 & 0.0758 \\
GCov (1 factor) & 0 & 0 & 0 & 0.2 & 5.3 & 19.5 & \textbf{75} & 0.0762 & 0.0943 \\
GCov (0 factor) & 0 & 0 & 0 & 0.8 & 13 & 34 & \textbf{52.2} & 0.1261 &  0.1598 \\

\bottomrule
\end{tabular*}

\vspace{3pt}
\begin{minipage}{\textwidth}
\footnotesize
\emph{Notes:} The data-generating process is a six-dimensional purely causal VARX model (\textit{$n1=6$}) from Equation~\ref{eq:varx2} with eigenvalues $(0.85,\,0.75,\,0.6,\,0.5,\,0.4,\,0.3)$ and the first three empirical factors from \cite{McCrackenNg2021FREDQD}, with the following factor loadings $\Gamma$ = [1.8 0.9 0.7; 0.9 1.8 0.6; 0.7 0.6 1.8]. Results are based on 1,000 Monte Carlo replications with sample size $T=1000$.
\end{minipage}
\end{table}

The results are reported in Table~\ref{tab:causal_varX2}. The main Monte Carlo statistic of interest is the empirical distribution of $ n_1$ across replications. The results provide clear evidence that factor filtering plays a role in removing spurious noncausal detection. When three factors are included, the estimator concentrates most of its mass at the true causal dimension ($n_1 = 6$ in 70.1\% of replications), with very limited dispersion toward lower dimensions. As the number of included factors decreases, the distribution shifts away from the true value: with two factors, the mass at $ n_1 = 6$ declines to 67.9\%, and with one factor to 68.6\%, while the frequency of underestimation increases. In the extreme case without factor filtering, the estimator exhibits only 52.2\% of replications correctly identifying the true dimension and a large share incorrectly selecting lower dimensions (e.g., 34\% at $ n_1 = 5$ and 13\% at $ n_1 = 4$). This deterioration is also reflected in the error metrics, with MAE deteriorating from 0.0589 (three factors) to 0.1261 (no factors). 

Overall, these findings support the hypothesis that noncausality in small-information VAR models can arise from omitted common factors rather than genuinely noncausal dynamics. Properly accounting for common information can restore correct causal classification. In the next section, we investigate whether similar effects arise in an empirical monetary policy VAR.

\section{Empirical application to monetary policy}
\label{sec:empirical}

This section revisits the benchmark small-scale monetary policy VAR model of \citet{stockwatson2001}, widely used in applied macroeconometrics, and re-estimates the system allowing for mixed causal-noncausal dynamics. Our objective is to assess whether the standard fundamental VAR model masks nonfundamental components arising from informational limitations, which may manifest as noncausal dynamics in the estimated system. We therefore re-estimate the benchmark model under this broader specification and examine the role of alternative policy rules and augmented information sets.

\subsection{Baseline VAR and policy-rules}
\label{sec:BaselineVAR}
We apply the proposed methodology with the tuning parameters of the GCov found in Section~\ref{sec:montecarlo} to the small-scale VAR approach originally employed in \citet{stockwatson2001}, which remains a standard reference for the analysis of U.S. monetary policy (see, e.g., \citealp{sophocles2021_ZLB_SW}). The vector of endogenous variables is
\begin{equation}
Y_t = (\pi_t, u_t, R_t)^{\prime},
\end{equation}
where $\pi_t$ denotes inflation, $u_t$ the unemployment rate, and $R_t$ the short-term nominal interest rate. Inflation is measured as the annualized quarterly growth rate of the GDP price index $\pi_t = 400 \log(P_t/P_{t-1})$,
unemployment corresponds to the civilian unemployment rate, and the interest rate is proxied by the federal funds rate. All variables are observed at a quarterly frequency.\footnote{Although we are aware that such a simple VAR model could be adapted, for instance by including additional variables such as oil prices or excess bond premium (see \citealp{gertler2015}), our interest in the \citet{stockwatson2001} simple model lies in the fact that the authors carefully describe and document their data and detail the way they transform them. 
}

\begin{figure}[h]
    \centering
    \includegraphics[width=0.8\linewidth]{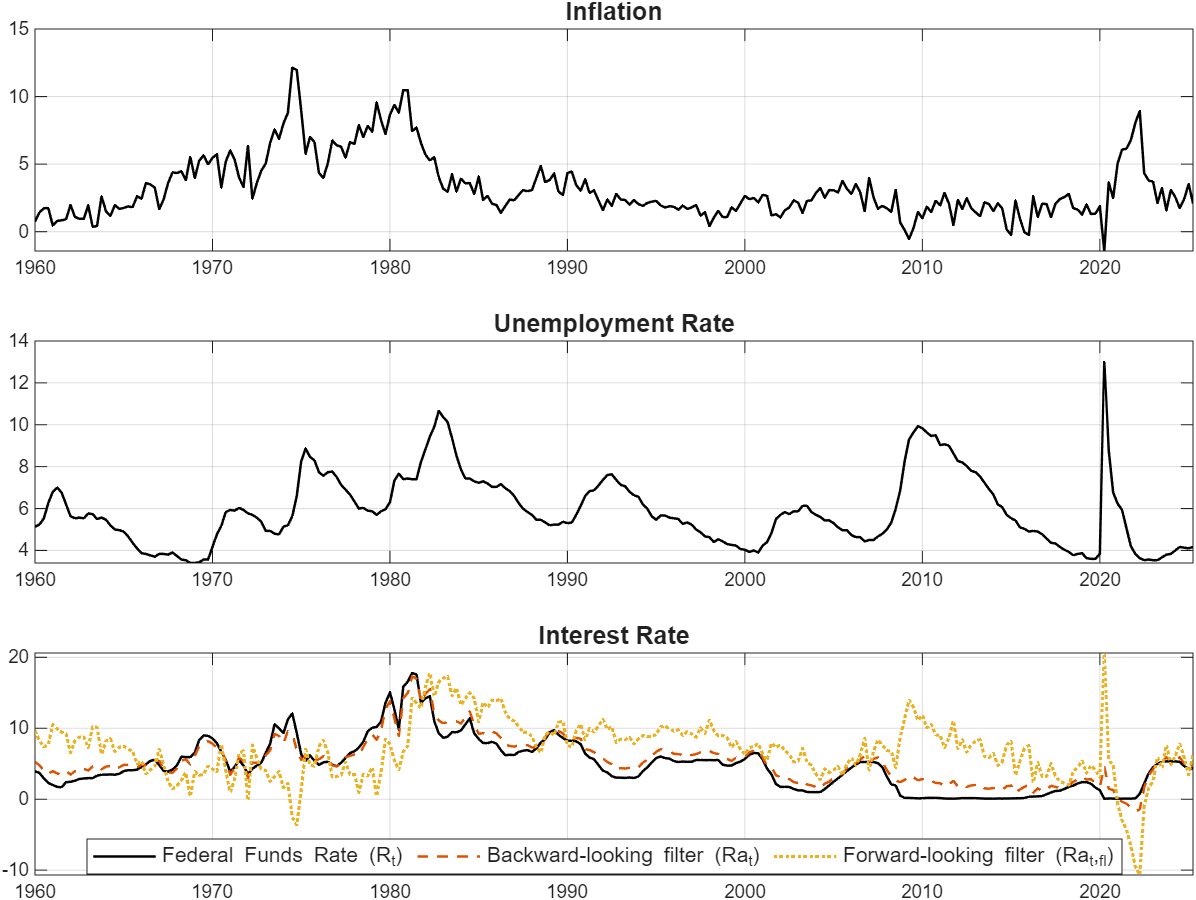}
    \caption{U.S. Inflation, Unemployment rate and Interest rate (filtered with backward and forward Taylor rules) from 1960:I to 2025:II.}
    \label{fig:sw_series}
\end{figure}

\citet{stockwatson2001} estimate a VAR with four lags over the sample 1960:I-2000:IV and study monetary policy identification by replacing the interest-rate equation with policy-rule restrictions. Keeping the same benchmark, we allow for mixed causal-noncausal dynamics in the VAR. We also extend the sample period to 2025:II in a second analysis. A lag-order selection based on information criteria (specifically the BIC) leads us to consider a VAR(2) specification instead of the VAR(4) estimated by \citet{stockwatson2001}.\footnote{Impulse response functions are very similar across the causal VAR(4) and VAR(2) specifications, in terms of both persistence and magnitude (see \citealp{stockwatson2001} for the four-lag framework).} For everything else we consider exactly the same data transformations as in Stock and Watson. The three series are represented in Figure~\ref{fig:sw_series}. From the results of the ADF tests, we might have chosen different transformations than the ones considered in \citet{stockwatson2001}. On the period 1960:I-2000:IV, the p-values of the ADF test with a constant only are 0.052, 0.060 and 0.077 for $\pi_t$, $u_t$ and $R_t$, respectively. Over the period 1960:I-2025:IV those p-values are respectively 0.025, 0.011 and 0.22. We see that at a 10\% significance level we could have taken the same transformation. For the federal funds rate, however, a different transformation could have been used over the full sample. Nonetheless, we stick to the way the authors transform the series in their paper. Residual diagnostics indicate non-Gaussianity, making the GCov estimator well-suited in this setting.

To relate the results to policy-rule representations of systematic monetary policy behavior, we filter the observed policy rate using Taylor-rule specifications following \citet{stockwatson2001}. As a benchmark, the original \cite{Taylor1993} rule describes systematic policy as a response to deviations of inflation and economic activity from their target values,
\begin{equation}
R_t = r^{\ast} + 1.5(\pi_t - \pi^{\ast}) + 0.5(y_t - y^{\ast}) + \varepsilon_t^{MP},
\label{eq:taylorRule}
\end{equation}
where $r^{\ast}$ denotes the equilibrium real interest rate, $\pi^{\ast}$ the inflation target, and $y_t - y^{\ast}$ the output gap. In practice, \citet{stockwatson2001} adapt the original Taylor rule by replacing the output gap using Okun’s law. In particular, the output-gap term $0.5(y_t - y^{\ast})$ in (\ref{eq:taylorRule}) is approximated by $-1.25(u_t - u^{\ast})$, yielding a policy rule expressed in terms of inflation and unemployment. Building on this framework, we consider two policy-rule specifications that replicate exactly the identification strategy of \citet{stockwatson2001}.

The first specification is a backward-looking Taylor rule, in which the policy rate responds to inflation and unemployment, together with their lagged values and lagged interest rates, with coefficients scaled to reflect a four-quarter horizon. Using this rule, we construct a filtered interest rate series, denoted $Ra_t$, by removing the systematic policy component implied by lagged inflation, unemployment, and interest rates from the observed policy rate $R_t$. The resulting residual captures deviations of the policy rate from the backward-looking Taylor rule and, following \citet{stockwatson2001}, is interpreted as a monetary policy shock.

The second specification is a forward-looking Taylor rule, in which policy responds to forecasts of inflation and unemployment four quarters ahead. In line with \citet{stockwatson2001}, these expectations are obtained from four-quarter-ahead forecasts computed from a reduced-form VAR(4).\footnote{While the GCov estimation is conducted on a VAR(2) selected by information criteria, the VAR(4) is used only to construct the expectations entering the policy rule, ensuring comparability with the original Stock-Watson framework.} We construct an alternative filtered interest rate series, denoted $Ra_t^{fl}$, by removing the systematic component implied by this expectation-based policy rule. Here, it is important to note that although referred to as a forward-looking rule, the specification does not involve noncausal dependence on future realizations of the data. Expectations are generated from VAR-based forecasts using the information set at time t. Further information regarding this filtering procedure can be found in the online appendix (Appendix A). 

Comparing results across $R_t$, $Ra_t$, and $Ra_t^{fl}$ (see Figure~\ref{fig:sw_series}, panel 3) allows us to assess how alternative representations of systematic monetary policy behavior affect the identification of noncausal dynamics in the VAR. Figure~\ref{fig:sw_series} shows that the interest rate filtered using the forward-looking Taylor rule exhibits more pronounced and persistent fluctuations during periods of high macroeconomic stress, in particular during the 2008 financial crisis and the post-COVID-19 episode. Relative to the baseline and backward-looking specifications, the forward-looking rule generates sharper deviations consistent with an increased role for expectations in policy behavior.

The VAR(2) parameters are estimated using GCov, with OLS estimates used as starting values for the optimization and nonlinear autocovariances computed up to lag $H=6$. Table~\ref{tab_SW_GCov} reports the estimated number of noncausal dimensions obtained from GCov estimation of a VAR(2) specification under alternative monetary policy rules for the original sample (1960:I-2000:IV). Across all specifications, the results indicate the presence of at least one noncausal dimension in the data.\footnote{Indeed it is not clear whether we have a root that is not different from 1 on the unit circle. This is probably due to the interest rate series. Appendix B (see online materials) shows how the presence of near-unit root in the true DGP affects the GCov performance in a Monte Carlo setting.} Importantly, this finding holds both for the original Stock-Watson sample period and for the extended sample through 2025 (see online Appendix C, panel A), providing empirical support for the relevance of mixed causal-noncausal dynamics in a canonical monetary policy VAR. This suggests that the econometrician’s information set may be incomplete, with omitted information manifesting as noncausal dynamics in small-scale systems. 
\begin{table}[h]
\centering
\footnotesize
\caption{GCov estimation on Stock and Watson (2001) original data}
\label{tab_SW_GCov}
\begin{tabular*}{\textwidth}{@{\extracolsep{\fill}} l l c}
\toprule
Specification 
& Eigenvalues 
& \begin{tabular}[c]{@{}c@{}}Noncausal \\ dim.($n_2$)\end{tabular} \\
\midrule

Baseline VAR $(\pi_t, u_t, R_t)$
& $\{\,0.2504\!\pm\!0.5614i,\;0.8770\!\pm\!0.9546i,\;0.3343,\;2.6804\,\}$
& 3 \\[2pt]

Backward-looking rule $(\pi_t, u_t, Ra_t)$
& $\{\,0.0843\!\pm\!1.0089i,\;0.8258\!\pm\!0.2753i,\;1.0182\!\pm\!1.5829i\,\}$
& 4 \\[2pt]

Forward-looking rule $(\pi_t, u_t, Ra_t^{fl})$
& $\{\,-1.7246,\;0.6029\!\pm\!0.4778i,\;1.1209\!\pm\!1.0169i,\;3.9810\,\}$
& 4 \\[6pt]
\bottomrule
\end{tabular*}

\vspace{2pt}
\footnotesize
\emph{Notes:} GCov with transformation set
$\{e_t,\ e_t^2,\ e_t^3,\ e_{ti}e_{tj}\}$ and nonlinear autocovariances up to $H=6$.\\ 
\end{table}

\subsection{Factor-filtered VARs}\label{sec:factorextraction}
Having established the presence of noncausal dynamics in the Stock-Watson model, we next examine whether these results are driven by omitted common macroeconomic information. To address this question, we filter the observable macroeconomic variables $Y_t$ to remove broad fluctuations captured by factors extracted from the FRED-QD dataset following \citet{McCrackenNg2021FREDQD}. The VAR is then re-estimated using the same GCov methodology.

\subsubsection{Database and factor extraction}
\label{sec:factor_extraction}

We use the quarterly FRED-QD database of \citet{McCrackenNg2021FREDQD}, an updated and expanded version of the large macroeconomic dataset of \citet{StockWatson2012a,StockWatson2012b}. The latest vintage contains $T=265$ quarterly observations on $N=245$ series spanning a broad set of U.S. macroeconomic indicators.\footnote{The series cover 14 categories, including real activity, labor markets, prices, interest rates, credit, housing, exchange rates, and financial variables.} Following the procedure of \citet{McCrackenNg2021FREDQD} (see Online Appendix D.1), we extract eight common factors. Since the explanatory power of the factors declines sharply beyond the first three, our analysis focuses on these leading factors (similarly to \citealp{FORNI2014fAVAR,BEAUDRY2019221} who also focus on the first three macro factors in a different SVAR setting).

Economic interpretation follows \citet{McCrackenNg2021FREDQD}. Factor~1 captures real economic activity and loads heavily on employment, industrial production, and NIPA series. Factor~2 reflects anticipated economic conditions, with strong loadings on capacity utilization, housing permits, and price-related indicators. Factor~3 primarily captures price dynamics and monetary aggregates. Online Appendix Table~3 reports the series with the largest marginal explanatory power for each factor, confirming these interpretations, while Figure~2 (Appendix) plots the estimated factors. Estimation of univariate mixed causal-noncausal models indicates that Factor~1 is purely noncausal, whereas Factors~2 and~3 are purely causal in both the original Stock-Watson sample and the extended sample.

\subsubsection{The Forni Gambetti test for informational sufficiency}
\label{sec:sufficiency}

A VAR can only recover the structural shocks if its variables span the information set of economic agents.  \cite{FORNI2014fAVAR} show that, under their assumed factor structure, a small VAR is informationally sufficient if and only if the latent factors do not Granger-cause the variables included in the VAR.  This result leads to a simple two-step diagnostic. First, the unobserved states are proxied by the leading principal components of a large macroeconomic panel. Second, one tests whether these components have incremental predictive power for the VAR variables, over and above the variables' own past. Informational sufficiency is rejected when the null of no Granger causality is rejected.\footnote{It is worth noting that informational sufficiency may hold for a particular structural shock even when it fails for the full set of structural shocks, a distinction known as partial fundamentalness (see \cite{FORNI2014fAVAR}). We leave this issue beyond the scope of the present study.}

We implement this diagnostic for $Y_t=(\pi_t,u_t,R_t)'$ over 1960:I-2000:IV, using the leading factors extracted from the FRED-QD panel introduced earlier. Following \cite{FORNI2014fAVAR}, we rely on the multivariate out-of-sample regression test of \cite{GELPERCroux2007_grange}. It compares the recursive one-step-ahead forecasts of a restricted VAR(2), containing only the own lags of $Y_t$, with those of a VAR(2) augmented with lags of the first $P$ principal components. The last 80 quarters (1981:I-2000:IV) are used for forecast evaluation, and $p$-values are obtained from a residual bootstrap with 500 replications. Table~\ref{tab:sufficiency} reports the results for $P=1,\dots,8$.

\begin{table}[htbp]
\centering
\caption{Out-of-sample Granger-causality test of informational sufficiency of $Y_t=(\pi_t,u_t,R_t)'$}
\label{tab:sufficiency}
\resizebox{\textwidth}{!}{%
\begin{tabular}{lcccccccc}
\toprule
 & $P=1$ & $P=2$ & $P=3$ & $P=4$ & $P=5$ & $P=6$ & $P=7$ & $P=8$ \\
\midrule
Reg statistic       & 15.09 & 52.62 & 62.85 & 55.41 & 26.90 & 31.15 & 33.60 & 32.29 \\
$p$-value           & 0.252 & $<0.002$ & $<0.002$ & $<0.002$ & 0.008 & $<0.002$ & $<0.002$ & 0.004 \\
\bottomrule
\end{tabular}}
\begin{minipage}{0.95\textwidth}

\vspace{2pt}
\footnotesize
\emph{Notes:} The null hypothesis is that the first $P$ principal components do not Granger-cause $Y_t$ (informational sufficiency of the baseline VAR). Reg is the regression statistic of \cite{GELPERCroux2007_grange}. Forecast evaluation sample: 1981:I-2000:IV (80 one-step-ahead forecasts, expanding window). $p$-values are based on the bootstrap distribution of the test statistic (500 draws).
\end{minipage}
\end{table}

Under our baseline VAR(2), the first factor alone does not carry significant predictive content for $Y_t$, but this does not indicate sufficiency: a small set of candidate predictors can fail to detect a deficiency that a richer set reveals. Once the second factor is added, the null of informational sufficiency is rejected at the 1\% level, and this rejection persists for every $P=3,\dots,8$. With four lags, the null is rejected already for $P=1$. The three-variable VAR is therefore informationally deficient, irrespective of the lag length. The test statistic relative to its critical value is largest for $P=3$, which supports filtering the system with up to $J=3$ factors, the number beyond which the explanatory power of the factors declines sharply. This makes it relevant to examine whether the noncausal dynamics detected in the baseline VAR reflect this informational gap.

\subsubsection{Filtering using macroeconomic factors}
This subsection examines how increasing the information set with common macroeconomic components affects the estimated dynamics of the VAR and its fundamentalness properties. We consider three alternative filtering schemes. Panel~A reports the baseline results from Section~\ref{sec:BaselineVAR}, while Panels~B-D report results obtained after filtering the variables using FRED-QD factors. Specifically, Panel~B filters the data using the first factor only ($F_{1,t}$), Panel~C uses the first two factors ($F_{1,t}, F_{2,t}$), and Panel~D uses the first three factors ($F_{1,t}, F_{2,t}, F_{3,t}$). In all cases, only contemporaneous factors are included. We focus first on the original sample of \cite{stockwatson2001}, namely (1960:I-2000:IV).\footnote{Factors are initially extracted using the full available FRED-QD sample (1959:Q2-2025:Q3) following \cite{McCrackenNg2021FREDQD}, and then restricted to match the \cite{stockwatson2001} sample period. Appendix D (online material) assesses the robustness of this approach by extracting factors only from the information set available at the time (1959:Q2-2000:Q4). The main conclusions remain unchanged, both for our analysis and \cite{FORNI2014fAVAR} information sufficiency test.}

The filtered VAR is then constructed using the vector $\tilde{Y}_t = (\tilde{\pi}_t,\tilde{u}_t,\tilde{R}_t)'$ and Equation~(\ref{eq:varxp}).
When the interest rate is replaced by the Taylor-rule-filtered series $Ra_t$ or $Ra_t^{fl}$, the same procedure is applied with $Y_t = (\pi_t,u_t,Ra_t)'$ or $Y_t = (\pi_t,u_t,Ra_t^{fl})'$, respectively. Using GCov, we estimate the resulting VAR(2) model for these filtered series under each factor specification ($J=1,2,3$) in order to evaluate how conditioning on an increasingly rich information set affects the degree of noncausality detected in the data.

The results, reported in Table~\ref{tab:gcov_factors_SW}, indicate that filtering by additional macroeconomic factors progressively reduces the number of noncausal dimensions across all three policy specifications.
While the unfiltered system (\textit{Panel~A}) and the specification filtered by the first factor only (\textit{Panel~B}) still exhibit noncausal components, the degree of noncausality drops once the first two factors are removed. In particular, for the Taylor-rule specifications based on $\tilde{Ra}_t$ and $\tilde{Ra}_t^{fl}$, no noncausal dimension is detected in \textit{Panels~C} and \textit{D}, indicating that purging common macroeconomic components is sufficient to recover a purely causal representation of the policy-rule VAR. This pattern suggests that part of the noncausal behavior identified in the unfiltered data is driven by aggregate macroeconomic forces captured by the common factors.\footnote{Results remain robust when including all eight extracted factors: in particular, the forward-looking Taylor-rule specification continues to display no noncausal dimension after filtering.
Also, it is worth noting that the VAR specifications in \textit{Panels~C} and \textit{D} satisfy the nonlinear serial dependence (NLSD) diagnostic test for i.i.d. residuals introduced by \cite{jasiak2025gcovportmanteau}, whereas the models in \textit{Panels~A} and \textit{B} fail this specification test, giving additional support for the relevance of the resulting purely causal representation.}

\begin{table}
\centering
\footnotesize
\caption{GCov estimation of VARs under alternative policy measures and factor filtering, Stock-Watson (2001) sample (1960:I-2000:IV)}
\label{tab:gcov_factors_SW}
\begin{tabular*}{\textwidth}{@{\extracolsep{\fill}} l l c}
\toprule
Specification 
& Eigenvalues 
& \begin{tabular}[c]{@{}c@{}}Noncausal \\ dim. ($n_2$)\end{tabular} \\
\midrule

\multicolumn{3}{l}{\textit{Panel A: No factor filtering}} \\[4pt]

Baseline VAR $(\pi_t, u_t, R_t)$
& $\{\,0.2504\!\pm\!0.5614i,\;0.8770\!\pm\!0.9546i,\;0.3343,\;2.6804\,\}$
& 3 \\[2pt]

Backward-looking rule $(\pi_t, u_t, Ra_t)$
& $\{\,0.0843\!\pm\!1.0089i,\;0.8258\!\pm\!0.2753i,\;1.0182\!\pm\!1.5829i\,\}$
& 4 \\[2pt]

Forward-looking rule $(\pi_t, u_t, Ra_t^{fl})$
& $\{\,-1.7246,\;0.6029\!\pm\!0.4778i,\;1.1209\!\pm\!1.0169i,\;3.9810\,\}$
& 4 \\[6pt]

\multicolumn{3}{l}{\textit{Panel B: Filtering by Factor 1}} \\[4pt]

Baseline VAR $(\tilde{\pi}_t, \tilde{u}_t, \tilde{R}_t)$
& $\{\,-0.1791,\;0.4172,\;0.9443,\;1.2321\!\pm\!0.6900i,\;1.2943\,\}$
& 3 \\[2pt]

Backward-looking rule $(\tilde{\pi}_t, \tilde{u}_t, \tilde{Ra}_t)$
& $\{\,-0.0684,\;0.4445,\;0.9251,\;0.8512\!\pm\!0.5501i,\;1.2300\,\}$
& 3 \\[2pt]

Forward-looking rule $(\tilde{\pi}_t, \tilde{u}_t, \tilde{Ra}_t^{fl})$
& $\{\,0.1074\!\pm\!0.2428i,\;0.9539,\;0.8920\!\pm\!0.5035i,\;1.3805\,\}$
& 3 \\[6pt]

\multicolumn{3}{l}{\textit{Panel C: Filtering by Factors 1-2}} \\[4pt]

Baseline VAR $(\tilde{\pi}_t, \tilde{u}_t, \tilde{R}_t)$
& $\{\,-0.0515,\;0.4932,\;0.9958\!\pm\!0.0564i,\;0.7805\!\pm\!0.9932i\,\}$
& 2 \\[2pt]

Backward-looking rule $(\tilde{\pi}_t, \tilde{u}_t, \tilde{Ra}_t)$
& $\{\,0.0715\!\pm\!0.1959i,\;0.9705\!\pm\!0.0678i,\;0.7884\!\pm\!0.5065i\,\}$
& 0 \\[2pt]

Forward-looking rule $(\tilde{\pi}_t, \tilde{u}_t, \tilde{Ra}_t^{fl})$
& $\{\,-0.1279,\;0.4167\!\pm\!0.4096i,\;0.6368,\;0.9962\!\pm\!0.0590i\,\}$

& 0 \\[6pt]

\multicolumn{3}{l}{\textit{Panel D: Filtering by Factors 1-3}} \\[4pt]

Baseline VAR $(\tilde{\pi}_t, \tilde{u}_t, \tilde{R}_t)$
& $\{\,0.1344\!\pm\!0.1288i,\;-0.3893,\;0.9681,\;1.0782\!\pm\!0.0837i\,\}$
& 2 \\[2pt]

Backward-looking rule $(\tilde{\pi}_t, \tilde{u}_t, \tilde{Ra}_t)$
& $\{\,0.1292\!\pm\!0.0689i,\;-0.5042,\;0.7262,\;0.9404\!\pm\!0.0954i\,\}$
& 0 \\[2pt]

Forward-looking rule $(\tilde{\pi}_t, \tilde{u}_t, \tilde{Ra}_t^{fl})$
& $\{\,0.1542,\;0.1410\!\pm\!0.3239i,\;0.8095,\;0.9861\!\pm\!0.1145i\,\}$
& 0 \\[2pt]
\bottomrule
\end{tabular*}

\vspace{2pt}
\footnotesize
\justifying
\emph{Notes:} Panel~A reports GCov estimates using unfiltered data from the original Stock-Watson (2001) sample. Panels~B-D report estimates based on variables filtered by contemporaneous FRED-QD factors. Nonlinear autocovariances are computed up to $H=6$, except for $\tilde{Ra}_t^{fl}$ of panel D which requires H=10 for convergence.
\end{table}

\subsubsection{Robustness Check}
The filtering step treats the factors as exogenous to $Y_t=(\pi_t,u_t,R_t)'$, which could fail if the FRED-QD panel contains series that are part of, or close substitutes for, the VAR variables. To assess this possibility, we first re-extract the factors after removing 51 series: the federal funds rate, the unemployment rate, the GDP price index and their mechanical decompositions, as well as close substitutes such as other interest rates and spreads, alternative price indices and monetary aggregates (see online appendix~D.2.2). The resulting factors remain closely aligned with the original ones, with absolute correlations of 0.99, 0.90 and 0.84 for the first three. The main findings are unchanged. Filtering with the first factor alone leaves three to four noncausal dimensions. Once the first two factors are removed, the noncausal dimension falls to zero in both Taylor-rule specifications, and to one in the baseline VAR. Therefore, our conclusions do not depend on the inclusion of series that may be too close to the VAR variables in the factor extraction. Additionally, \cite{Bernanke2005_FAVAR} estimate a FAVAR in which the federal funds rate $R_t$ and factors extracted from a large panel are modeled jointly, and since the raw principal components may load directly on $R_t$, they purge this contamination by regressing the components on this interest rate. Following the same procedure, we additionally test the robustness of our model by creating a subgroup of the FRED-QD panel that only includes slow-moving variables (201 variables), assumed to not react contemporaneously to Rt (see more details on appendix~D.2.2) and extract the factors ($\hat F^{\mathrm{slow}}_t$) on that panel, similarly as before. We then obtain our new factors, purged from Rt, as $\hat F^{\mathrm{purged}}_t=\hat F_t-\hat \beta R_t$, where $\hat F_t$ are the principal components of the full FRED-QD panel (245 series) and $\hat \beta R_t$ is obtained from the regression of $\hat F_t$ on the slow-moving factors $\hat F^{\mathrm{slow}}_t$ and $R_t$. As in the first exercise, filtering with the first two purged factors and estimating with GCov yields purely causal specifications for both Taylor-rule cases.

We now extend the analysis to the full sample (1960:I-2025:II). Detailed results are reported in the online appendix (Appendix C). In this extended sample, factor filtering does not fully eliminate the noncausal components detected in the baseline specification, even when all eight factors are included. This suggests that some noncausal dynamics remain and may stem from additional sources of omitted information not captured by the FRED-QD dataset, or from other forward-looking behavior that is difficult to measure using standard macroeconomic indicators. Alternatively, the difference between the original and extended samples may reflect structural breaks or regime changes happening in the economy after 2001. This potential issue is already noted in \cite{stockwatson2001}, who point out that time variation in monetary policy rules may lead to misspecification of constant-parameter VAR models.\footnote{Time-varying parameter noncausal autoregressive models have already been developed by \citet{LanneLuoto2017} for the specific case of U.S. inflation. Extending this approach to noncausal VAR models is left for future research.} 
Although we recover evidence of noncausality in the full sample, some have argued that nonfundamentalness is not necessarily an issue in the recovery of unbiased impulse responses, especially when the VAR conveys large information (see, i.e., \citealp{BEAUDRY2019221} and \citealp{forni2018_nonfund_DSGE} for the case of technological news shocks).
Yet, to better understand the timing of the appearance of noncausal dynamics, we re-estimate the model using an expanding window starting in 2001, updated every quarter. The results indicate that noncausal dimensions begin to appear in the mid-2000s, following the 2008 financial crisis. Interestingly, we observe a further increase in the number of noncausal dimensions after the COVID-19 episode in late 2019-early 2020.\footnote{We use the full information set to estimate this expanding window (past and future) to capture the forward looking behavior of agents. More precisely, factors are extracted over the full sample 1960:Q1-2025:Q2, and backward- and forward-looking Taylor-rule components are estimated on the same sample. The model is then estimated using an expanding window with updated subsets of factors and interest-rate filters. This reconstruction allows us to approximate the information set available to agents and better capture forward-looking behavior.}

\begin{figure}[htbp]
    \centering

    \begin{subfigure}{0.49\linewidth}
        \centering
        \includegraphics[width=\linewidth]{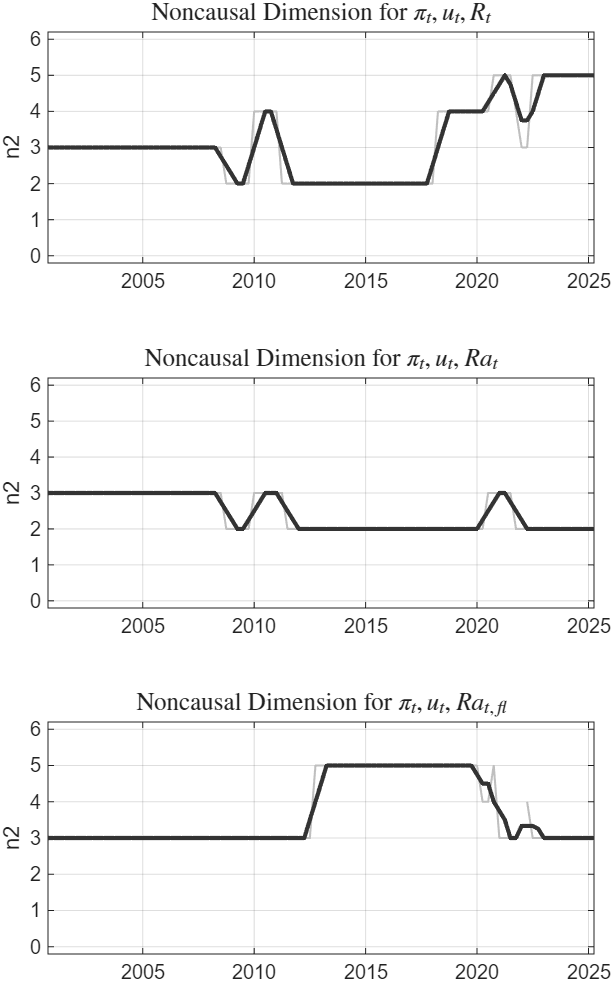}
        \caption{Unfiltered (J=0)}
        \label{rolling_unfiltered}
    \end{subfigure}
    \hfill
    \begin{subfigure}{0.49\linewidth}
        \centering
        \includegraphics[width=\linewidth]{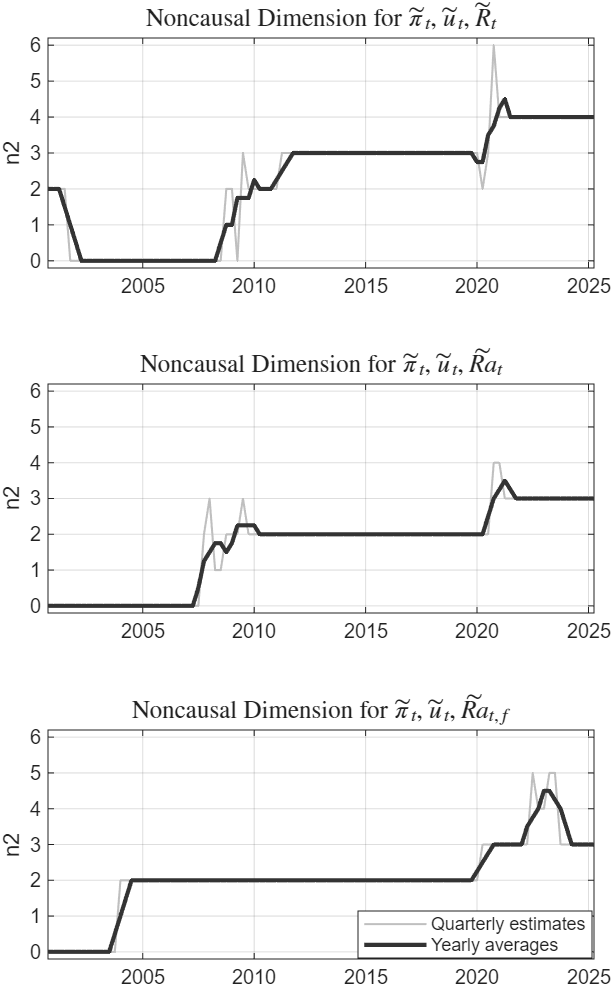}
        \caption{Filtered by Factors (J=3)}
        \label{rolling_filtered}
    \end{subfigure}

    \caption{Expanding-window estimates of the noncausal dimension based on the GCov estimator. The left panel reports results for the unfiltered data and the right panel for the factor-filtered data (with J=3 factors). Estimation windows begin in 1960:Q1 and expand sequentially through 2025:Q2. The grey line plots quarterly estimates, and the black line plots the corresponding four-quarter moving average.}
    \label{fig:rolling_Window_RGCOV}
\end{figure}

\subsection{Impulse Response Functions}

In the presence of nonfundamental dynamics, standard VAR inversion-based impulse response functions are not necessarily economically interpretable, since a fundamental moving-average representation may not exist. We therefore restrict the IRF analysis to specifications that become purely causal after factor filtering, for which a fundamental MA representation is well defined and standard recursive identification is valid. As shown in Table~\ref{tab:gcov_factors_SW}, once the variables are filtered using at least the first two macroeconomic factors, the VAR(2) estimated with the GCov procedure on systems including the Taylor-rule residuals $\tilde{Ra}_t$ or $\tilde{Ra}_t^{fl}$ no longer exhibits noncausal dimensions. These factor-filtered specifications can thus be treated as standard causal VARs, allowing us to compute impulse response functions following the procedure in \citet{stockwatson2001}.

The IRF analysis is conducted on the ordered factor-filtered vector $\tilde{Y}_t$. In the policy-rule specifications, $\tilde{Y}_t=(\tilde{Ra}_t,\tilde{\pi}_t,\tilde{u}_t)'$ or $\tilde{Y}_t=(\tilde{Ra}_t^{fl},\tilde{\pi}_t,\tilde{u}_t)'$, where $\tilde{Ra}_t$ (resp.\ $\tilde{Ra}_t^{fl}$) denotes the deviation of the policy rate from the backward-looking (resp.\ forward-looking) Taylor rule described in Section~\ref{sec:BaselineVAR}. In the recursive benchmark without policy-rule restrictions, the ordering is $\tilde{Y}_t=(\tilde{\pi}_t,\tilde{u}_t,\tilde{R}_t)'$. Structural shocks are identified via a recursive (Cholesky) orthogonalization of the reduced-form innovations, which imposes a lower-triangular contemporaneous impact matrix. Under the policy-rule ordering, this restriction is coherent with the Taylor-rule construction: contemporaneous inflation and unemployment innovations may enter the policy residual, while macroeconomic variables do not respond contemporaneously  to the identified monetary policy shock.

\begin{figure}[htbp]
    \centering
        \includegraphics[width=\linewidth]{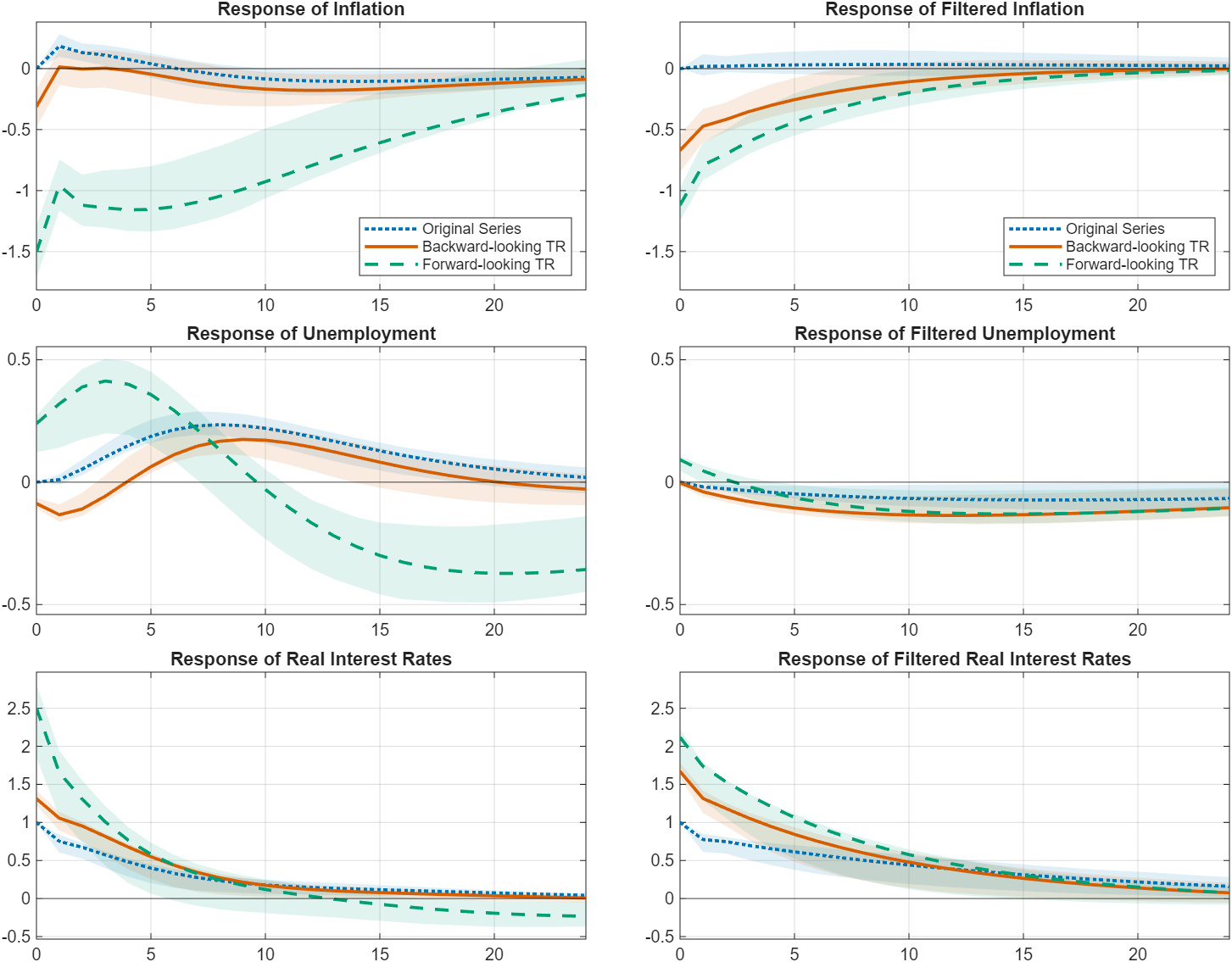}
    \caption{Comparison of impulse response functions to a contractionary monetary policy shock: unfiltered (left) and filtered with 3 factors (right), sample (1960:I-2000:IV). \\
    \textit{Note}: The impulse responses are computed exactly as in \citet{stockwatson2001}. The confidence bands are an addition to the original study, following  the standard 66\% bootstrap percentile bands (point estimate $\pm$ one standard error), based on 1,000 replications.
    }
    \label{fig:irf_comparison}
\end{figure}

Impulse responses are then obtained by iterating the implied moving-average representation of the estimated VAR(2). Following \citet{stockwatson2001}, shocks are scaled to correspond to a one-percentage-point innovation in the real policy rate, and responses are reported over a 24-quarter horizon. Figure~\ref{fig:irf_comparison} reports the IRFs for the unfiltered and factor-filtered specifications, and their confidence bands. We observe that using a Taylor-rule filter resolves the price puzzle and leads to a stronger disinflationary effect of monetary policy, both in the original and in the factor-filtered specifications, with the effect even more pronounced in the latter case. Although this result echoes earlier findings, it is consistent with the evidence in \citet{Bernanke2005_FAVAR}, who show that augmenting the information set with a large number of macroeconomic variables can eliminate the price puzzle. Also, once the data are filtered using macroeconomic factors, restrictive monetary policy shocks are also associated with a smaller, barely significant response of unemployment (see similar results in the review of \citealp{RAMEY201671}). Overall, these results suggest that enriching the econometric information set may affect the estimated transmission of monetary policy shocks, leading to a disinflationary response and suggesting, in our sample, a more limited economic slowdown through smaller estimated effects on real activity such as unemployment.

We conclude this section with a comment on the identification strategy. We deliberately keep the identification strategy of \citet{stockwatson2001} to ensure comparability with the original study. Nevertheless, we are aware that more recent approaches, such as sign restrictions \citep{Uligh2005_SR} and external instruments \citep{gertler2015, StockWatson2018_IV}, among others, are now standard and provide more robust identifications. Additionally, previous studies have already shown that FAVARs can attenuate the price puzzle \citep[e.g.,][]{Bernanke2005_FAVAR}. Our contribution instead lies in revisiting a seminal paper to show that nonfundamentalness should not be blindly assumed, but rather tested. In doing so, we aim to bring additional empirical evidence of nonfundamentalness to the ongoing debate, and to raise awareness of tools, including mixed causal-noncausal models, that can capture dynamics overlooked by small conventional VARs. For this reason, we simply replicate the original methodology and report the IRFs for additional discussion purposes.

\newpage
\section{Conclusions}
\label{sec:Conclusions}

This paper studies whether nonfundamentalness in small-scale VARs can be explained by omitted macroeconomic information. To this end, we introduce a mixed causal-noncausal VARX framework that combines noncausal autoregressive dynamics with common factors. Monte Carlo evidence supports the ability of the proposed framework to account for nonfundamentalness (captured by noncausal roots of the VAR parameter) arising from omitted common information and indicates that the GCov estimator performs well in this setting.

Empirically, we revisit the structural VAR framework of \citet{stockwatson2001}, using the same variables (inflation, unemployment, and the interest rate under alternative Taylor-rule specifications) and the same sample period. We first estimate a VAR(2) that allows for mixed causal-noncausal dynamics and document robust evidence of noncausal components across all interest-rate specifications. We then investigate whether these dynamics reflect omitted aggregate information available to economic agents but not to the econometrician. To do so, we filter the variables using macroeconomic factors extracted from a large information set and re-estimate the system. Once the first factors are removed, the noncausal components disappear in the Taylor-rule specifications, suggesting that the previously detected noncausality was largely driven by omitted common information. Because the factor-filtered systems become purely causal, impulse response functions can be computed within a standard fundamental framework. We show that this adjustment further mitigates the price puzzle and strengthens the disinflationary effects of monetary policy shocks.

Interestingly, when the sample is extended through 2025, noncausal dynamics remain present regardless of the Taylor-rule specification or factor filtering. This finding calls for further research on how to construct economically interpretable impulse response functions in mixed causal-noncausal models, for instance in line with the nonlinear innovation-filtering approaches developed by \citet{gourieroux2026nonlinear}.

\newpage
\bibliographystyle{chicago}
\bibliography{referencesme2.bib}

\end{document}